# ASCEND: Personal AI Agents for Autonomous Scientific Computing Across HPC Clusters and GPU Workstations

*J. Paul Liu[1], Uthpala Herath[2], Andrew Petersen[3]*

*[1.] AI Hub for Science, College of Sciences, NC State University, Raleigh, NC; email: jpliu@ncsu.edu*

*[2.] Research Computing and Support Services, Office of Information Technology, Duke University, Durham, NC; email: uthpala.herath@duke.edu*

*[3.] Office of Information Technology, NC State University, Raleigh, NC; email: aapeters@ncsu.edu*

## Abstract

Traditional scientific computing requires researchers to translate computational intent into environment configuration, resource requests, and executable jobs, then diagnose failures from scheduler state and application logs. We present **ASCEND (Autonomous Scientific Computing Engine and Novel Discovery)**, an AI-powered agent system, operated through a command-line or a local web chat interface, that supports several placements of the agent and, in the arrangement used for every case reported here, runs it on the researcher's own laptop, reaching Slurm-managed clusters and a GPU workstation over a multiplexed authenticated connection, with site-specific execution policies checked by locally executed tools. The language model agent (Claude Code or Codex, selected by the researcher at each launch) is hosted remotely and proposes actions but holds no credentials. No facility-scale service is required: an account on each resource is sufficient, and the public installer lets users link additional Slurm clusters or workstations of their own. Every launcher offers a per-session choice of agent runtime. We report four recorded cases: 1) The agent closed a failure-recovery loop on a planted tensor-device fault, submitting, diagnosing, repairing and resubmitting with job-level artifacts preserved. 2) It reproduced the published evaluation of a weather-forecasting model from the author's released forecasts, recovering an evaluation protocol the paper does not fully state, resolving an ambiguity in that protocol empirically, and agreeing with the published curves to 2.1 percent on z500 and 2.4 percent on t850, while identifying a unit discrepancy in the paper's prose and an initialization-field discrepancy, possibly of regridding origin, in its released data. 3) It parallelized a released 12,693-line geophysical solver under a requirement of bit-for-bit identity of the exported outputs with the serial build, reducing wall-clock runtime from about twelve hours to about two. 4) That requirement exposed two instances of undefined behaviour in the published solver, one of which silently erased cells at the flow front and changed simulated results; both were confirmed with independent analysis tools, repaired, and reported to the upstream authors. These cases were operated through ordinary scheduler commands rather than the policy-checked submission path; a separate, pre-specified evaluation of that optional validator found that it rejected 29 of 30 constructed violations and held the remaining one (GPU node hour) for approval, while denying 3 of 14 legitimate requests. The system's autonomy was exercised under author supervision using the Claude Code runtime, and agent-level transcripts were not retained, so human-intervention rates and the attribution of individual repair decisions are not independently verifiable; a controlled end-to-end recovery benchmark remains outstanding and requires an isolated cluster.

## 1 Introduction

Executing a scientific computation on shared infrastructure requires decisions about software environments, storage, resource allocation, and scheduler state. A researcher who delegates code generation to a language model may still need to relay errors, submit jobs, and decide when a failed run can be retried. Existing workflow systems already automate dependency management and execution, and programmatic HPC interfaces support remote access. The remaining question is how an agent can adapt an authorized workflow when assumptions about its environment fail [1, 2].

Throughout, the language model is hosted remotely and accessed over the network via API: it proposes and inspects actions but holds no credentials or execution access to site infrastructure or the scheduler. The interface is agent-runtime-agnostic: each launch presents a choice between two commercially available agent CLIs, Anthropic's Claude Code and OpenAI's Codex. Because site procedure resides in reviewed instruction documents and locally executed tools rather than in any one model's recollection, the runtime is interchangeable at the point of launch; all cases reported here were run under Claude Code. Our **Autonomous Scientific Computing Engine and Novel Discovery-ASCEND** system addresses this operational layer through a shared set of command-line tools and site adapters. The supplied design covers Hazel, the NCShare HPC cluster, and the Hurricane GPU workstation, representing dedicated submission-node, laptop-proxy, and direct-execution arrangements; Hazel supports both the submission-node and the laptop-proxy forms. The intended workflow allows a scientist to specify a goal while the agent prepares jobs, observes failures, and proposes or performs bounded repairs. Resource escalation remains a human decision when it exceeds the initial authorization [3].

This paper argues that **an AI agent, governed by an explicit execution policy and reaching remote resources over an ordinary authenticated connection, is a practical operating layer for autonomous scientific computing available to an individual researcher.** The arrangement requires no facility-scale deployment and no service the institution must build: a laptop, an account on each resource, and a multiplexed connection are sufficient. Section 4 reports four recorded cases in which the system carried scientific work through that loop end to end, reproducing a published result and recovering its unstated evaluation protocol, parallelizing a released solver under a bitwise-reproducibility constraint, and identifying two defects in that solver which the constraint exposed and which have been reported upstream. Section 5 reports the executed portion of a pre-specified evaluation protocol for the policy layer itself. We do not claim a new language model, a general repair algorithm, or a measured improvement in the rate of scientific discovery; the claim is that the operating layer works, on real shared infrastructure, for a single researcher.

# 2 Related works

## 2.1 Agents for research computing

ORNL's cross-facility architecture combines agents, facility services, and provenance, with a demonstration linking simulated manufacturing components to HPC [4]. Purdue's Hello Computer describes institutional agent context, MCP interfaces, and deployment guidance [5]. Like ASCEND, Hello Computer operates local-first tools through the researcher's existing SSH credentials with no hosted service, so the overlap is substantive: ASCEND's distinction rests on its typed execution boundary, routing and recorded validation cases rather than on the absence of a facility-scale service alone. These establish important precedents for scientific orchestration and site-aware access. Our ASCEND system concentrates on the execution and recovery behavior of researcher-controlled jobs across several operating arrangements.

HPC-AutoResearch combines phased execution, repair, containers, and memory [6]. Foam-Agent supplies domain-specific simulation repair and HPC submission [7], while VibeCodeHPC targets iterative HPC code optimization [8]. Consequently, automated repair and retained context are existing ideas. ASCEND's prospective contribution is their integration with explicit site-policy validation, evaluated independently of the model's knowledge of local procedures.

Descriptive Execution evaluates natural-language HPC workflows [9], and EngiAI measures completion of successive remote-training stages [10]. These motivate validation of terminal scientific outputs rather than submission alone. Hierarchical Server Architecture addresses resource negotiation [11], while Academy and RASER provide federated-agent and resilient-runtime infrastructure [12, 13]. ASCEND should complement these layers rather than claim to replace them.

## 2.2 Interfaces and authority

SWE-agent and RepairAgent establish precedents for task-specific interfaces and autonomous repair [14, 15]. InfrastructureSentinel applies policy controls to infrastructure agents [16], and an independent dtu-hpc-mcp project describes restricted operations with resource ceilings [17]. These works make the enforcement boundary, rather than tool naming or transport protocol, the relevant comparison. HPC-specific security research further distinguishes valid credentials from authorized task behavior [18].

Several systems now place a chat or browser interface in front of agent-operated HPC. IrisAI, deployed institutionally at Memorial Sloan Kettering, serves a browser chat with MCP-wrapped Slurm tools, skills and policy controls for a cancer-research user community [19]; it is a facility-side service, where ASCEND is a local, single-user arrangement with no server-side component. Operon, an agent IDE for bioinformatics, runs headless agent sessions on cluster nodes over authenticated SSH with multi-factor login, streamed tool events and session resumption [20], an arrangement close to ASCEND's remote placements. FermiLink couples coding agents with domain skills into a unified framework for autonomous multidomain scientific simulations [21], overlapping ASCEND's scientific-workflow scope. Browser access to HPC is itself established practice through Open OnDemand [22], and Chat AI connects a web service to Slurm through restricted SSH, though to serve models on HPC rather than to let agents operate scientific workloads [23]. These systems separate three concerns — web access, model serving and autonomous execution — and the browser-to-scheduler connection of section 3.6 accordingly has precedents; ASCEND's contribution rests on the typed execution boundary, the multi-site deployment arrangements, and the recorded scientific cases of section 4.

Broader systems such as The AI Scientist and its successor target research generation [24, 25], while ReAct and Reflexion provide foundations for tool interaction and feedback memory [26, 27]. ASCEND's present scope is narrower: reliable operation of an already specified computation, with scientific correctness determined by an external validation procedure.

# 3 System design

## 3.1 Shared tools and site adapters

The project identifies *ascend* as the coordinating interface, *hpcrun* as the Slurm-aware job preparation and submission tool, *hpcrepro* as an environment and failure-diagnosis component, and fetch-paper as a literature-access utility. Common logic resides in a shared core; adapters capture site-specific launch paths, environment setup, and operational rules, distilled from each site's published user documentation (the NCShare user guide [28] and NC State's Hazel HPC documentation [29]). The coordinating agent's reasoning runs as a cloud service via the selected provider's API (Anthropic's for Claude Code, OpenAI's for Codex); the model receives task context and tool outputs but holds no credentials or execution access to site infrastructure, and every action it proposes is executed locally by the tools below, which is also where site policy is enforced. Runtime selection is part of the launch sequence: each launcher prints its site banner and then asks which agent CLI will drive the session, showing the installed version of each or an offer to install the one that is missing. The choice is per-session and deliberately not remembered, so selecting a runtime remains a conscious act rather than a default; both runtimes read the same seeded instruction file and reach the resources through the same tools and multiplexed connection. The revision evaluated in section 5 is hpcrun 0.5.0, with the harness revision and tool checksums recorded in the evaluation package. Table 1 states what is implemented today, the evidence behind each component, and the remaining limitations.

```
$ hpcrun --site hazel --gpus 1 --time 02:00:00 train.py
$ hpcrepro --job 123456 --diagnose
```

*Listing 1. Illustrative tool invocations; the exact schemas accompany the code release.*

Site knowledge itself is packaged as skills: versioned instruction documents, one per job class per resource, that record the site's launch paths, environment and module conventions, and known failure signatures. A skill is authored once from a shared template, reviewed by a person, and installed alongside the harness; the agent loads the matching skill when a task begins, so local procedure comes from the reviewed document rather than from the model's recollection of site conventions. Updating a procedure means editing the skill, not re-prompting the model, and the design keeps one skill per job class to avoid overlapping guidance. Skills are installed and versioned alongside the harness, and the evaluation package records the revision, so evaluated behavior can be attributed to a specific skill revision. For resources beyond the institutional presets, a site-agnostic variant of the skill ships in place of a reviewed per-site document (section 3.6): it directs the agent to the site's own documentation installed alongside it and to live probes of the scheduler and environment, and the agent's first session distills these into a dated site profile that subsequent sessions read as the local procedure, open to the same human review and versioning.

*Table 1. Implementation status: what is implemented today, the evidence behind it, and what remains.*

| Component | Implemented behavior | Evidence | Remaining limitation |
|---|---|---|---|
| hpcrun 0.5.0 | Slurm-aware job preparation, typed resource requests, deterministic pre-submission policy validator with budget threshold | Deployed validator probed in 5.2 (29 rejected, 1 held of 30) | Checks are static and lexical; not a mandatory gateway and not exercised by the section 4 case, which submitted through plain sbatch (3.2) |
| ascend launchers + ascend-all router | Per-arrangement launchers installed by the 3.6 packages; availability probe, recommendation, confirmed dispatch | Recorded routing session (Figure 3); author-run end-to-end installs (3.6) | Routing quality bounded by the 5.1 feasibility analysis; no independent-user study |
| hpcrepro | Environment and failure diagnosis, installed with the harness | Ships in every install package | Not on the execution path of the recorded section 4 case (plain sbatch on record) |
| Site skills | Versioned per-site instruction documents, one per job class, installed with the harness | Install packages; section 3.1 | Skill-revision attribution recorded only in the evaluation package |
| Multiplexed connection layer | One interactive login; 8-hour non-interactive reuse; setup scripts create aliases and warm the link | Author-run installs on both cluster paths (3.6) | Session expiration and reconnect behavior not measured; duplicate-submission handling unspecified |
| Action contract (Listing 2) | Proposed; hpcrun validates specification fields and one per-request budget threshold today | Budget hold observed in 5.2 | Full contract and cumulative (across-request) budget accounting not implemented |
| Provenance / memory store | Lesson store distinguishing unverified hypotheses from run-supported lessons | Section 3.5 | Memory benefit unevaluated (T16 held out); complete provenance schema not established |
| fetch-paper | Full-text literature retrieval via published-source resolution | Section 3.5 | No literature-to-discovery claim |
| Agent-runtime selection | Per-session choice of Claude Code or Codex in every launcher, with install-status display and an installer offer for a missing CLI | Author-run launches on both laptop-driven arrangements (2026-09-30) | All section 4 cases and the section 5 evaluation ran under Claude Code; no cross-runtime comparison |

Figure 1 shows how the shared tools connect agent decisions to the four deployment arrangements.

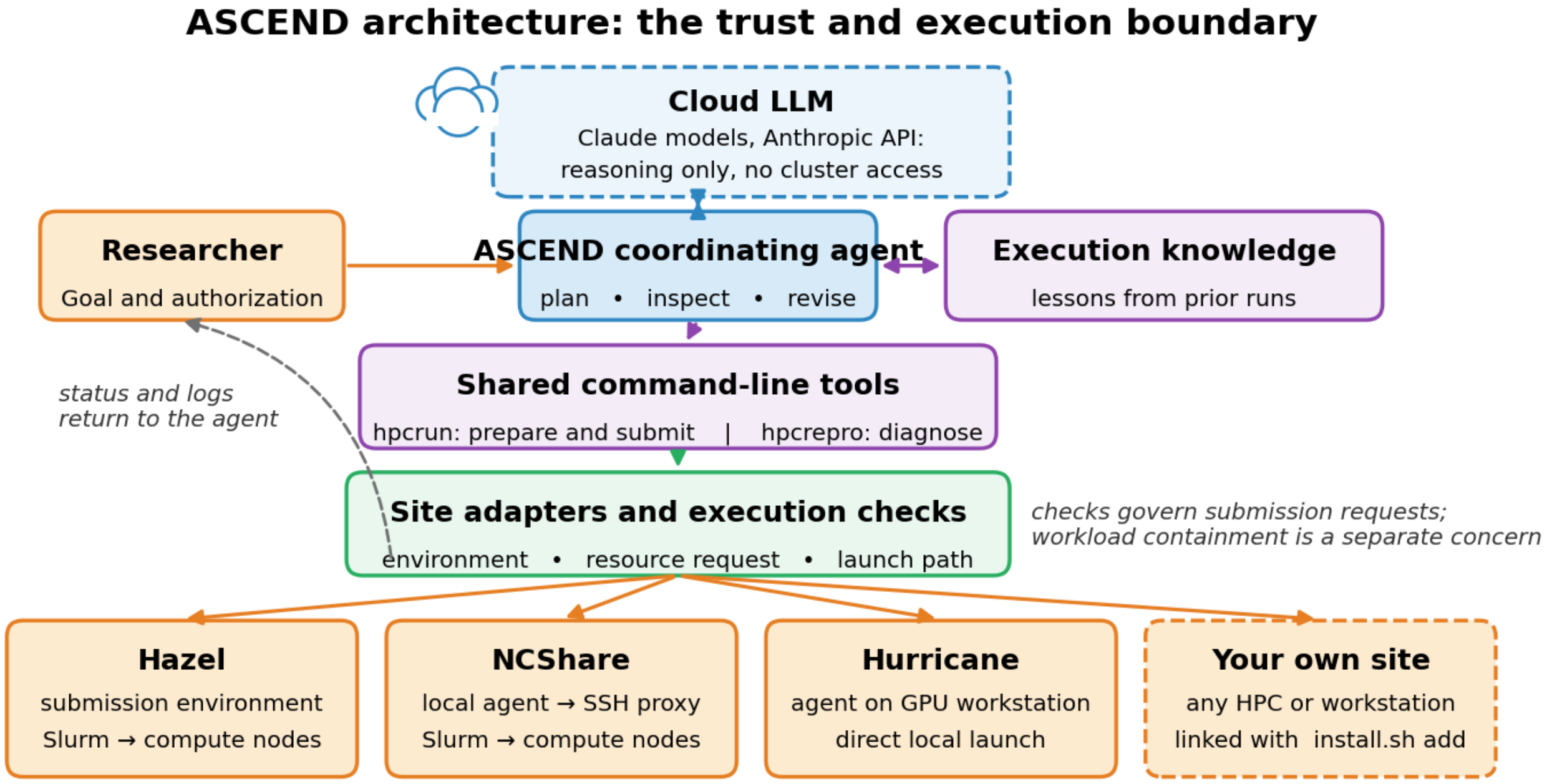

Figure 1. ASCEND architecture: the trust and execution boundary. A shared tool layer connects the coordinating agent to site-specific execution paths. Arrows indicate logical interaction. The diagram also shows the agent's connection to the selected cloud language model it reasons with; the drawing shows the Claude configuration

used for every recorded case, with Codex reached the same way when selected. Deployment breadth and enforcement completeness remain to be verified.

*Table 2. Deployment arrangements across the supported environments (users can link to their own systems).*

| Environment in the design | Agent placement | Execution path | Site responsibility |
|---|---|---|---|
| Hazel | Researcher's local machine or a dedicated submission environment (VCL) | SSH proxy or Native Slurm access | Scheduler policy and environment setup |
| NCShare | Researcher's local machine | SSH proxy to Slurm | Proxy handling and explicit resource requests |
| Hurricane | GPU workstation | Direct local launch | Local availability and execution limits |
| User-linked site (added via install.sh add) | Researcher's local machine, the linked cluster's login node, or the linked workstation | Multiplexed SSH; Slurm via hpcrun on a cluster, direct local launch on a workstation | Site profile distilled from the site's own documentation and a live probe; owning account's site policy applies |

Table 2 summarizes deployment arrangements reported in this paper. It does not establish that every feature has been tested in every environment. Hardware inventories, storage quotas, and retention periods are recorded in the dated deployment configuration in the evaluation package rather than as general architectural properties.

### 3.2 Execution boundary

The design routes execution through typed tools intended to validate resource requests and site rules before launch. Such validation can reject malformed or disallowed requests, but a typed interface does not by itself constrain every effect of submitted code. In the implementation as deployed, the answers are as follows. The agent retains ordinary shell access under the researcher's account on whichever machine it runs, and the scheduler remains directly reachable there: the recorded case in section 4 submitted through a plain sbatch invocation. hpcrun is therefore the policy-checked path, not a mandatory gateway or a sandbox; its validation governs requests routed through it, policy files are ordinary files owned by the researcher, and the reachable paths and credentials are exactly those of the researcher's account. Two distinct mechanisms therefore govern the agent and should not be conflated: site rules the agent is instructed to follow, which are advisory and depend on the agent honouring them, and checks mechanically enforced on requests routed through the validator, which are not. The cases of section 4 relied on the first: each submitted through ordinary scheduler commands under author supervision, and none was routed through the validator, whose behaviour is measured separately in section 5.2. Static lexical checks over a job specification must not be read as containment of arbitrary submitted programs, and the validator results in section 5.2 are scoped accordingly.

For evaluation, we propose an explicit action contract containing the operation, project root, target site, resource request, and allowed execution budget. A trusted validator would accept an action only when its schema is valid, its project and operation are authorized, its resources satisfy site constraints, and its cumulative use remains within the approved budget. This contract is a proposed specification to check against the implementation, not a claim that all fields or enforcement mechanisms already exist.

```
{
"operation": "submit",
"project_root": "~/projects/tensor-demo",
"site": "hazel",
"resources": { "gpus": 1, "time": "02:00:00", "mem": "32G" },
"budget": { "retries": 2, "gpu_hours": 8.0 }
}
```

*Listing 2. Proposed action-contract fields (illustrative); a trusted validator accepts an action only when every field satisfies the checks above.*

Scientific programs executed as batch jobs retain capabilities under the user's account unless separately contained. Therefore, submission-policy checks and workload isolation require separate descriptions and tests. The agent must also treat logs and retrieved documents as data, since otherwise a diagnostic channel can redirect subsequent actions.

### 3.3 Job lifecycle and recovery

The reported control loop prepares a job, submits it, observes its termination status and logs, diagnoses a failure, edits the relevant computation, and resubmits. A useful formalization distinguishes prepared, submitted, queued, running, failed, validating, and completed states. "Completed" should require both successful execution and a task-specific correctness check. These states are a proposed reporting model; their persistence and implementation remain to be documented.

Bounded recovery requires explicit retry limits and stopping conditions. Repeated failure, exhausted resources, an unavailable environment, or a proposed change to scientific intent should return control to the researcher. An agent must not obtain success by dropping required calculations, weakening an error tolerance, or omitting failed outputs. The evaluator should hold validation criteria outside the agent's editable workspace.

Remote operation begins with connection establishment. In the described design, the researcher authenticates interactively once, entering a password and completing a two-factor confirmation, to open a persistent multiplexed SSH connection using OpenSSH connection sharing (a control master with an extended persistence window). Subsequent tool invocations reuse this control channel non-interactively, so job submission, status polling, and log retrieval do not repeat the interactive exchange. This SSH channel is unrelated to the agent's connection to its language model, which is a separate, ordinary outbound connection to the selected provider's API (Anthropic or OpenAI) and carries no cluster credentials. This first step is what makes the laptop-proxy arrangement practical: without multiplexing, each command would require a fresh two-factor login, which is incompatible with an automated control loop. Setup scripts assist the warm-up by opening the authentication session and polling the control socket until the link is live. Multiplexing is an availability mechanism rather than an authorization mechanism; the shared connection carries the researcher's ordinary credentials and confers no additional privilege.

Figure 2 formalizes the recovery loop and separates execution success from scientific correctness.

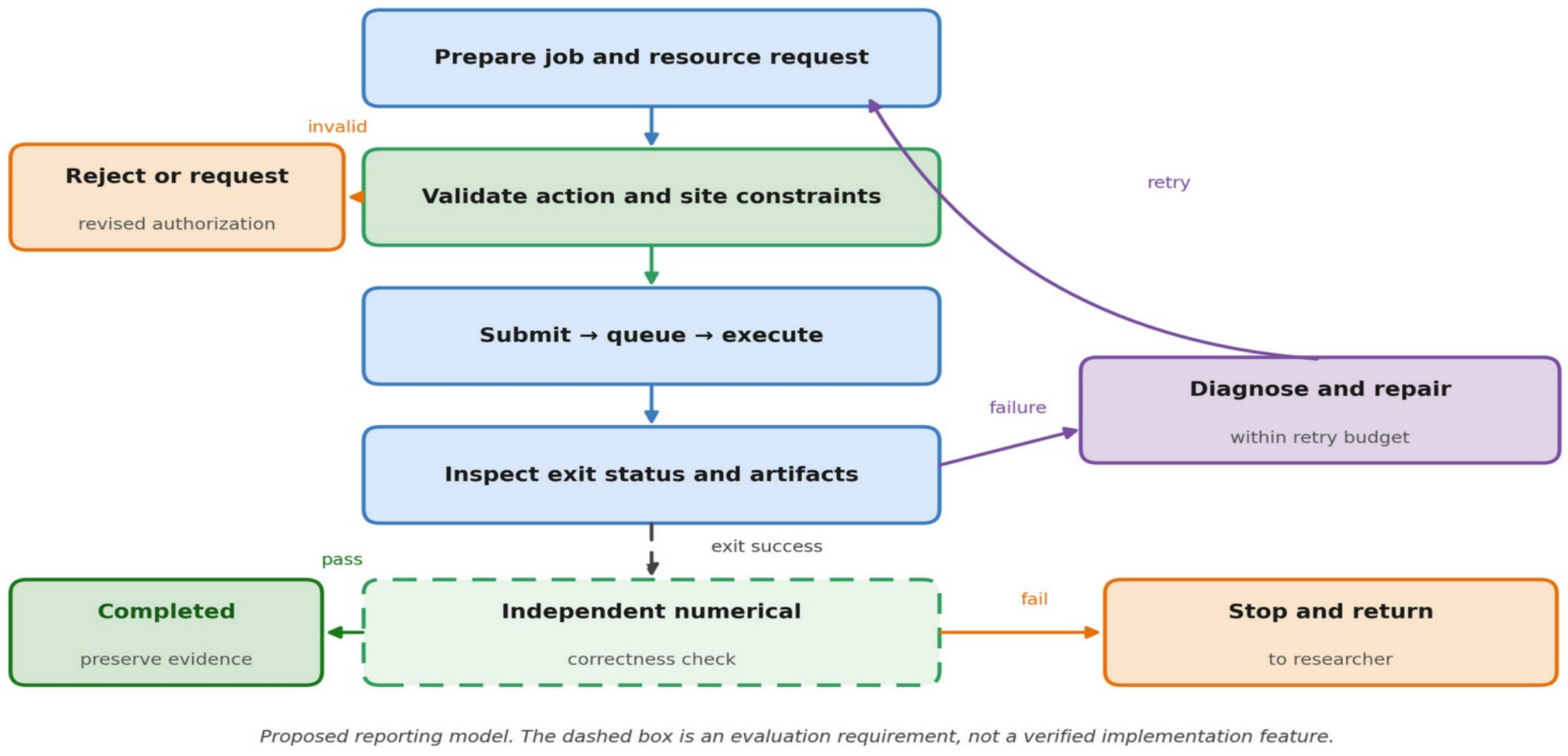

Figure 2. Proposed execution and bounded-recovery flow. Invalid actions return for revision; recoverable failures may be retried within the authorized budget. Exhausted retries or changes beyond authorization return control to the researcher. The dashed correctness check is an evaluation requirement, not a verified implementation feature.

```
Host yourhpcname   #ncshare, hazel or hurricane
   HostName login.example.edu
   User your_user_name
   ControlMaster auto
   ControlPath ~/.ssh/sockets/%r@%h-%p
   ControlPersist 8h
   ServerAliveInterval 60
```

*Listing 3. OpenSSH multiplexing configuration for the laptop-proxy arrangement (illustrative). One interactive login opens the control socket; subsequent tool invocations reuse it.*

The design uses SSH connection reuse and batching to support remote operation. These are transport mechanisms. Session expiration and reconnect behavior must be measured separately, and payload encoding should not be described as a security check. Duplicate submissions after a lost launch acknowledgement are not currently deduplicated; handling this is left to future work.

### 3.4 Routing and escalation

The proposed *ascend-all* workflow probes resource availability, recommends a destination, presents parameters to the researcher, and dispatches after confirmation. The confirmation prompt also accepts a decline that returns to a fresh job description, so a researcher can re-query the same availability snapshot without restarting, and the Hazel destination dispatches to the laptop-driven launcher, falling back to the dedicated-node launcher where only that arrangement is installed. An escalation after GPU memory exhaustion is illustrated in the design material and should be treated as design behavior until a corresponding execution trace is recorded.

An escalation is not necessarily live migration. Staging code and restarting on a new host, restarting from a checkpoint, and transferring an active process have different requirements. In particular, multiple GPUs do not automatically combine their memory for an application. The task must support a suitable parallel execution strategy, and output equivalence must be checked after any configuration change.

Figure 3 shows one such trace: a recorded *ascend-all* session, from the live resource snapshot through the job description, the model's recommendation and rationale, and the confirmation prompt at which the researcher accepts or overrides the destination. The routing verdict itself is obtained non-interactively through the locally installed Claude Code runtime; the launcher the router dispatches to then presents the per-session runtime choice of section 3.1.

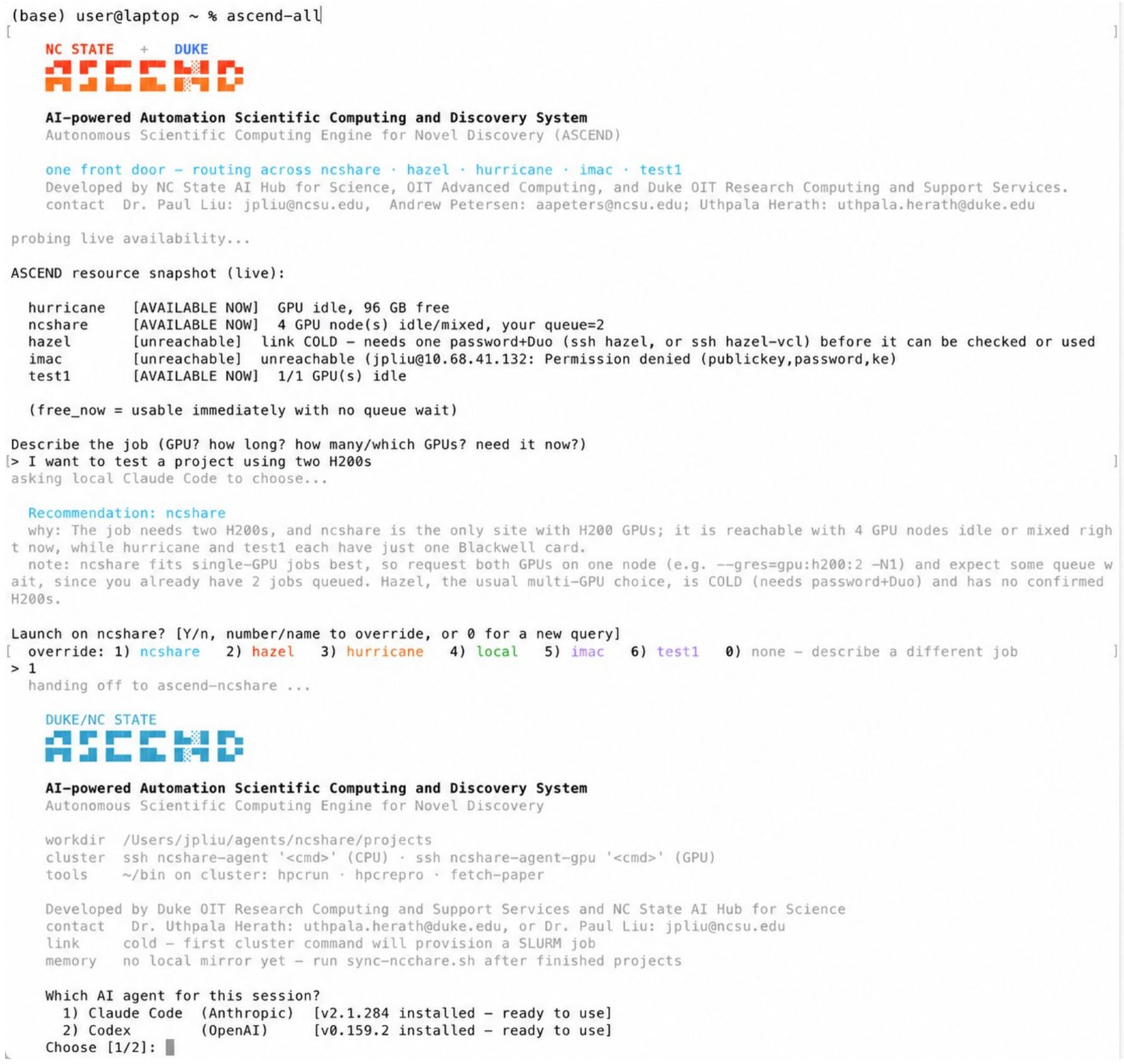

Figure 3. A recorded ascend-all routing session. The front door probes live availability across the three resources, takes a natural-language job description, obtains a recommendation and rationale from the cloud language model, invoked non-interactively through the locally installed agent runtime, and dispatches only after researcher confirmation. The session shown recommends the GPU workstation for a single-GPU reproduction task and notes the conditions under which it should move to a cluster instead.

### 3.5 Provenance and memory

ASCEND's reported memory design retains execution lessons and distinguishes unverified hypotheses from lessons supported by successful runs. The accompanying record should preserve the task specification, code revision, environment, site-policy version, scheduler identifiers, repairs, and correctness results. These fields define an artifact requirement for the study; the presentation does not establish a complete implemented provenance schema.

Literature retrieval feeds reproduction rather than standing on its own: retrieving a paper does not validate a computation derived from it, and the verification must come from an independent reference. Section 4.2 reports a case in which the two are combined, recovering an evaluation protocol from a published paper and checking the reproduced result against the published figure. Memory benefits are a separate question and require a controlled comparison that prevents prior exposure to evaluation tasks; none is claimed here.

### 3.6 Packaging and onboarding

The system is distributed as four self-contained packages, namely ascend-ncshare, ascend-hazel (the laptop-driven Hazel arrangement over the multiplexed login-node link), ascend-hazel-vcl, and ascend-universal, the last covering all arrangements including standalone workstations, together with a combined ascend-all bundle that ships the three laptop-driven arrangements and the routing front door behind one setup script. Each package is installed by a single setup script and preceded by a standalone install guide. The setup scripts also offer an optional install of the second agent runtime, so a new user can have both CLIs available from the first launch. The laptop-driven Hazel package removes the dependency on a reserved submission node and its session limits, and installs no agent runtime on the cluster; its login-node use is confined to job scheduling and environment builds, with all computation submitted through the scheduler. The scripts encapsulate the connection mechanics of section 3.3: they create the multiplexed SSH configuration and its socket directory, write any missing proxy aliases from the site's published proxy definition in the NCShare user guide [28] (userguide.ncshare.org/guides), warm the first authenticated link, and deploy the harness and skills to the resource. Agent-facing templates are tokenized and substituted with the installing user's identity at install time, so no other user's account names or paths ship in the packages.

Figure 4 shows the four arrangements as a user experiences them: which machine carries the agent, and what the multiplexed link connects.

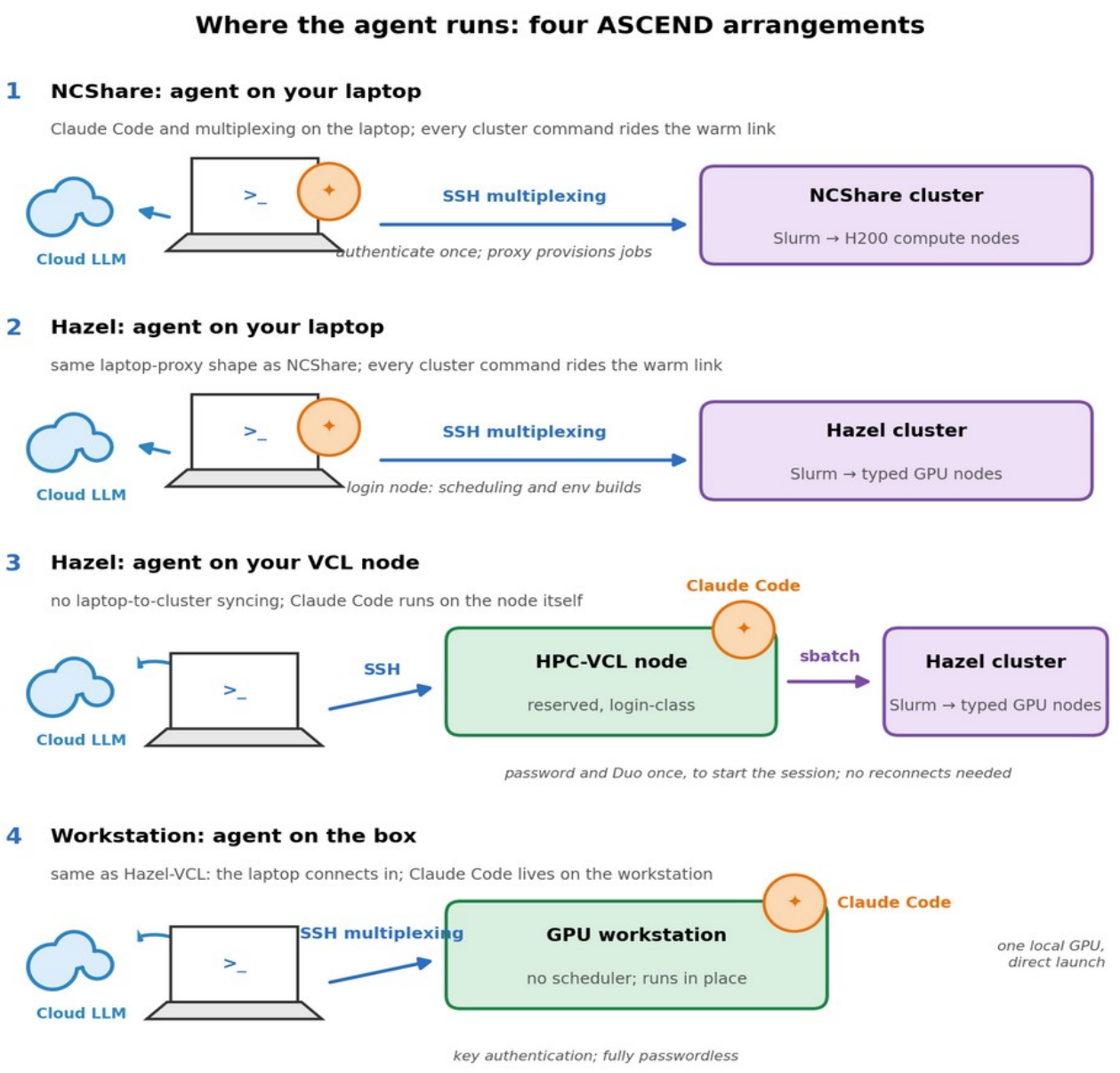

Figure 4. The ASCEND deployment arrangements. On NCShare, and on Hazel reached via the laptop, the agent runs on the laptop and every cluster command rides a multiplexed SSH link (on NCShare to a Slurm-allocated compute node, on Hazel to the login node, which is confined to scheduling and environment builds), authenticated once and reused across the many separate connections that follow. On Hazel reached via HPC-VCL, the agent

runs on the reserved node itself, so the laptop's SSH connection is used once to start that session rather than multiplexed for repeated reuse. A workstation follows the HPC-VCL shape with no scheduler. Wherever the agent runs also holds its own separate connection to the selected cloud language model (the Anthropic or OpenAI API), shown at the left of each row.

The configuration a new user supplies is deliberately small and site-shaped: an NCShare install asks only for the username behind an already-working SSH login; a Hazel VCL install asks for a Unity ID and the address of the user's reserved node; a workstation install asks for the machine's SSH login. The agent runtimes themselves are installed and authorized by the user with one command per machine; the launchers additionally offer that install whenever the runtime chosen for a session is missing, over the multiplexed connection when the agent runs on the remote machine.

Figure 5 shows the selection as the user encounters it, and the property that makes it safe: whichever runtime is chosen, the session converges on the same seeded instructions, the same typed tools, and the same multiplexed connection.

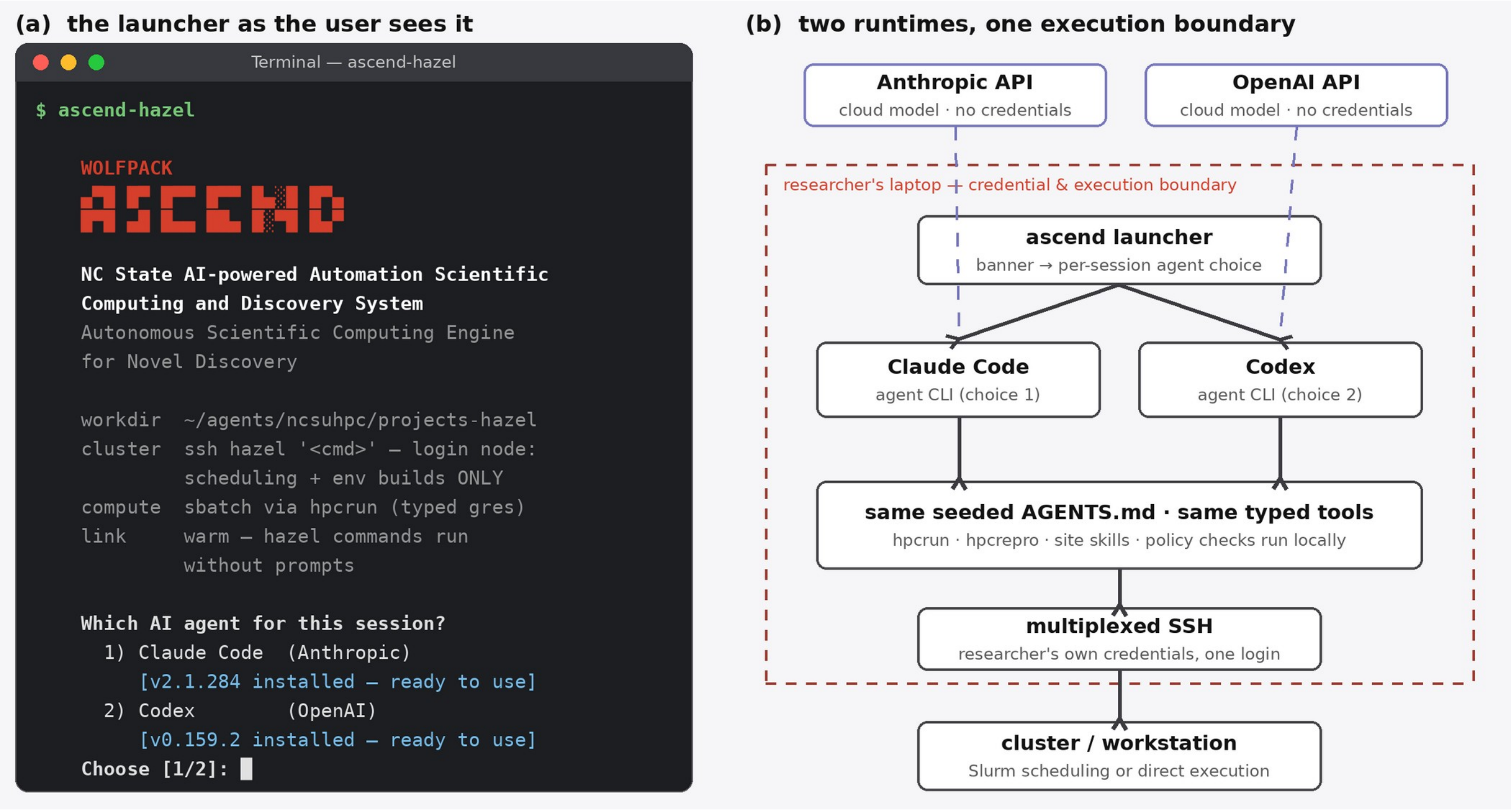

Figure 5. Per-session agent-runtime selection. (a) A recorded ascend-hazel launch: the site banner, then the runtime menu showing each CLI's installed version or an offer to install a missing one. (b) Both runtimes converge on the same instruction file, tool layer and multiplexed SSH link; the provider APIs sit outside the credential boundary and hold no credentials.

The same agent is also reachable from a browser. An optional web interface (install.sh web) starts a small local server on the researcher's computer — implemented with the Python standard library, bound to the local loopback address and protected by a per-start access token — and serves a chat page that drives the agent runtime in headless mode inside the selected site's project directory. Conversations persist across messages through the runtime's session-resumption mechanism, so a conversation begun in the browser can be continued from the terminal in the same directory, and vice versa. The page streams the agent's commentary and each executed command as the run proceeds, and files the agent writes into the project directory — reports, figures, logs — are listed under the reply and render inline (Figure 6). Because headless operation cannot present per-command approval prompts, the server runs the agent with pre-approved execution inside the same policy and credential boundary as a terminal session; the interface binds only to the researcher's machine and is single-user by design. The web interface currently drives the Claude Code runtime only.

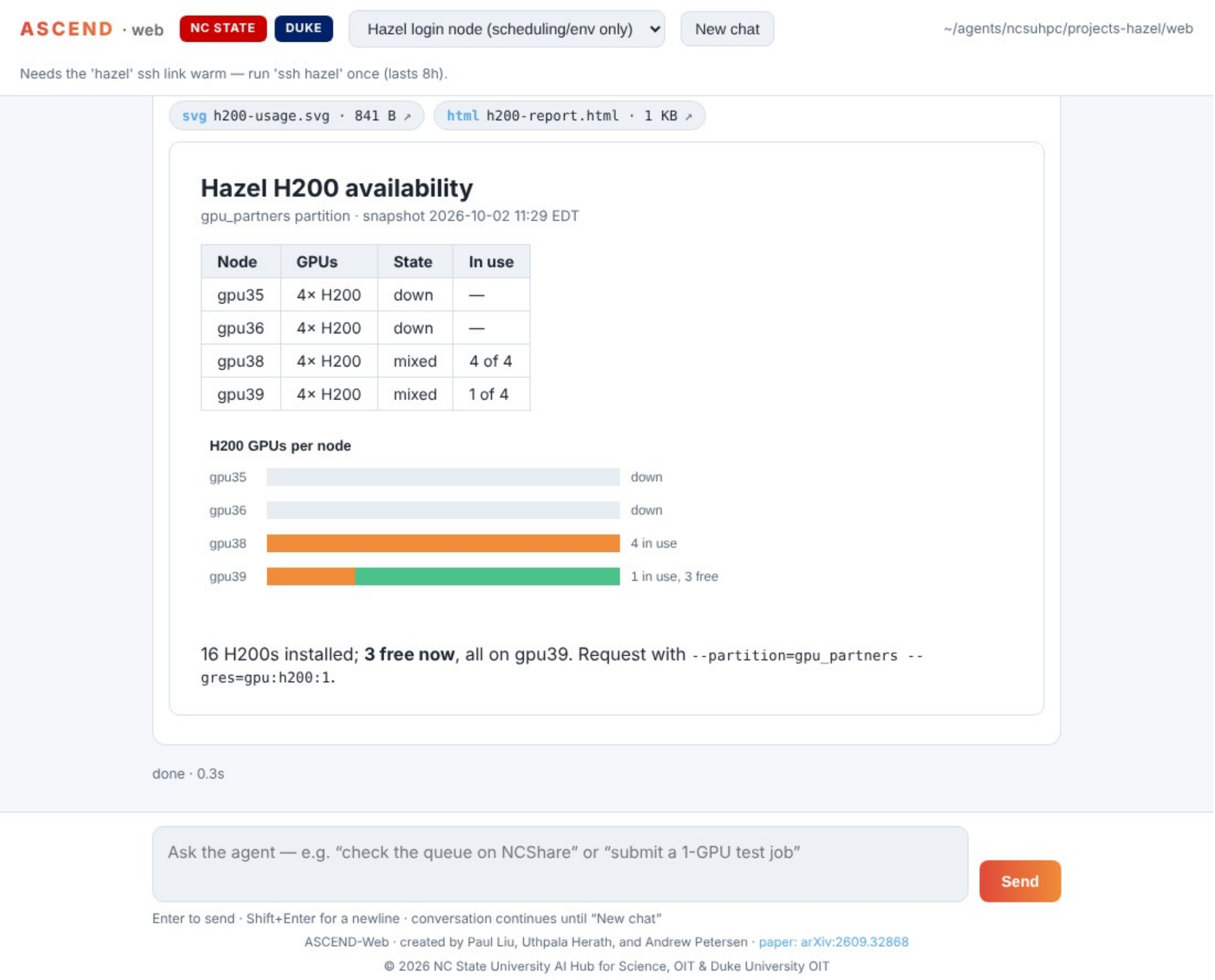

Figure 6. The ASCEND web interface. A browser chat page drives the same agent, seeded instructions, typed tools and multiplexed SSH boundary as the terminal launchers; each executed command streams into the page as the run proceeds, and files the agent writes into the project directory (here an HTML report) render inline under the reply.

The packaging is not limited to these presets. The public installer also provides a bring-your-own-site path (install.sh add) that links any Slurm cluster or workstation on which the user holds an account, such as a cluster at another campus or a departmental GPU server. An interactive wizard creates or reuses the multiplexed SSH alias of section 3.3, probes the remote host to classify it as a Slurm cluster, a GPU workstation without a scheduler, or a plain CPU host, asks whether the agent should run on the researcher's own computer (the recommended arrangement for shared login nodes) or on the linked machine itself, and deploys the shared harness together with the site-agnostic skill for that class. The wizard accepts the site's published user guide and policy documents, which are installed alongside the skill; as described in section 3.1, the agent's first session at the site distills them, with a live probe of the scheduler, quotas and module system, into a dated site profile that later sessions read first, so site-specific knowledge is generated from the site's own documentation rather than hand-written per site. Each linked site is registered with the routing front door of section 3.4, which thereafter probes its live availability and includes it among the candidate resources it recommends over. This path is implemented and distributed in the same repository; the validation reported below covers the preset arrangements, and no independent-campus deployment is yet claimed.

Both cluster paths were validated end-to-end by an author following the written guide on a fresh setup (2026-09-20). The walkthrough functioned as a test and exposed one first-run defect (launchers failed when their default working directory did not yet exist, a state no developer machine exhibits), which was corrected in all packages. Onboarding of users outside the author group has begun; no independent-user usability result is claimed.

# 4 Demonstrated capabilities

This section reports four recorded cases that together exercise the loop the system is built for: read the site rules, write or port the code, submit, read the logs, diagnose, repair, resubmit, and verify the result against an independent reference. The cases were run on a personal laptop driving NCShare over the multiplexed connection of section 3.3; no component of the agent runs on the cluster. All four cases were driven by the Claude Code runtime; the per-launch runtime selection postdates these recordings. In every case the work was agent-driven under author supervision: the author set the goal and approved resource requests, while the agent recovered protocols, wrote and ported code, submitted and monitored jobs, read failure logs, diagnosed causes, applied repairs, and surfaced the discrepancies reported below. Job identifiers, code, patches, logs and machine-readable results for all four cases are preserved in the evaluation package; agent-level session transcripts were not retained, which section 6 records as a limitation.

## 4.1 Closing the failure-recovery loop

The demonstrated case is a deliberately planted device-placement defect in a two-GPU PyTorch workload on NCShare (two NVIDIA H200 GPUs on the gpu-hp partition). The workload is a synthetic throughput soak (concurrent bfloat16 matrix multiplication followed by an NVLink peer-exchange phase) rather than a neural-network model. The planted fault allocates the exchange-phase accumulators on the CPU, so the job runs correctly for 150 seconds before raising the tensor-device mismatch, and diagnosis requires reading the failure log rather than the exit code alone. Two timestamps therefore appear in the record and measure different intervals: the fault triggers after roughly 150 seconds of correct execution (the “exit 1 at 150 s” in Figure 7), while the observed wall-clock failure at 2 minutes 47 seconds after the job's allocation began (the interval the scheduler logs support; queue wait is excluded) additionally includes startup and initialization.

In the recorded run (job 732794, 2026-09-16), the job failed 2 minutes 47 seconds after the job's allocation began (the interval the scheduler logs support; queue wait is excluded) with “RuntimeError: Expected all tensors to be on the same device, but found at least two devices, cuda:0 and cpu.” The repair on record is a single line, adding device=d to the accumulator allocation, and the resubmitted job (732795) started 29 seconds after the failure was logged and ran to completion. A secondary case recorded the previous day required two repair rounds: an unplanned submission-script fault (an unbound Slurm variable) followed by the planted device-count error, with a 78-second gap between the final failure and the clean rerun.

Figure 7 summarizes the recorded sequence and the evidence that accompanies it.

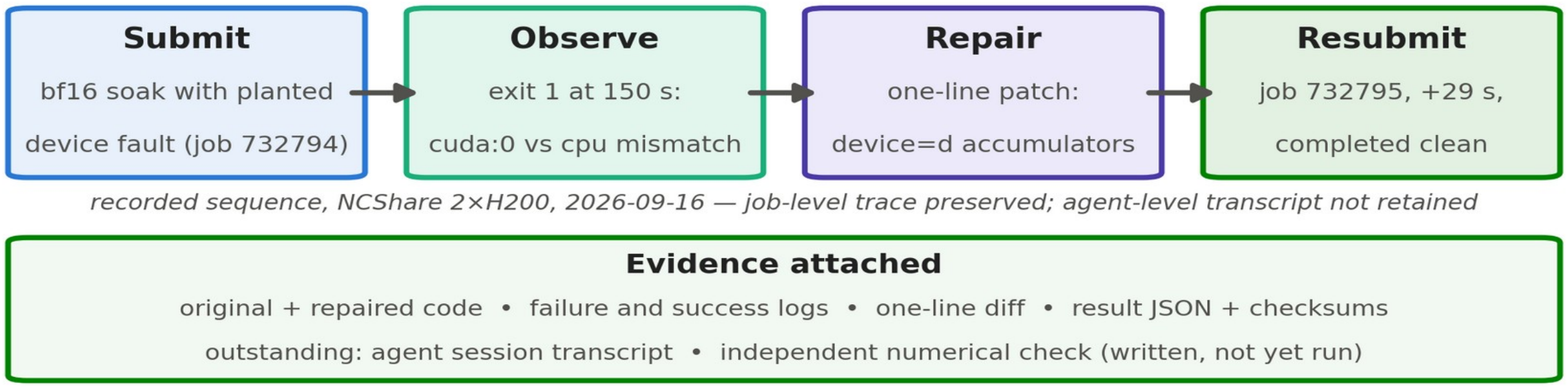

Figure 7. The recorded device-mismatch recovery case (jobs 732794 → 732795, NCShare, 2026-09-16). Job-level artifacts (code, patch, logs, and results) are preserved; the agent-level transcript and the independent numerical check remain outstanding.

Recorded performance characterizes the workload, not the agent: sustained bfloat16 throughput of 658.9 TFLOP/s per GPU (peak 30-second sample 718.8; first-half to second-half drift −0.38 percent) and 294.5 GB/s aggregate NVLink bandwidth in the primary run, and 756.8–777.8 TFLOP/s bfloat16 and 49.0–51.3 TFLOP/s fp32 in the secondary case’s sequential single-GPU benchmarks. The previously circulated figure of 252 TFLOP/s in fp16 matches no artifact on record and is withdrawn. Each configuration was run once, with host-side timing and no warmup in the primary case, on nodes shared with unrecorded neighboring GPU load, so these figures carry no uncertainty estimates. An arithmetic re-derivation of every figure from the recorded operation counts and elapsed times reproduced the compute numbers exactly and exposed a units error in the secondary case's bandwidth figures (GiB/s reported as GB/s, a 7.4 percent understatement), corrected in the evidence package.

The archive preserves job-level evidence for both cases: the original and repaired programs, failure and success logs, the one-line repair diffs, submission scripts, machine-readable results, and checksums. It does not preserve agent-level evidence: no transcript or tool-call log covers the 29-second diagnosis-and-repair interval, the submission path visible in the record is a plain sbatch invocation rather than the hpcrun or hpcrepro tools, and the zero-human-steps boundary therefore remains an author-reported observation that the artifacts neither confirm nor refute. An independent numerical check with seeded inputs, CUDA-event timing, and ten repetitions has been written but not yet executed; until it runs, the numerical correctness of the workload outputs is unverified. The case stands as a documented feasibility demonstration of the detect-diagnose-repair-resubmit loop, not a measurement of recovery performance.

The evidence package (code, patches, logs, machine-readable results, checksums, and the pending independent check) accompanies this paper; the outstanding items are the agent session transcript and the executed numerical check.

## 4.2 Reproducing a published result

The second case reproduces a published result the authors of this paper did not produce. The target is Keisler [30], Forecasting Global Weather with Graph Neural Networks. The author released the 2020 forecast set but not the training code or model weights, so the model cannot be retrained from the released artifacts; what can be reproduced exactly is the evaluation of the released forecasts, and that is what was done, on NCShare on 15 September 2026.

The evaluation protocol is not stated completely in the paper, and recovering it was most of the work. From the paper's own section 4.5 and its Figure 7 the agent reconstructed a 1 degree grid, Gaussian smoothing at FWHM 1.5 degrees to approximate a T120 truncation, an extratropics-only mask at latitude 20 degrees and poleward, all 732 initializations of calendar year 2020, a latitude-weighted mean squared error computed per forecast and averaged over initializations before taking the square root, and a 0.25 day leftward shift of the plotted curves. One step is genuinely ambiguous: the paper does not say whether the smoothing is applied to the ERA5 truth as well as to the forecast. Rather than guess, the agent computed all three variants and selected the one matching the published figure. This is protocol reconstruction rather than independent validation of the convention: the same figure supplies both the selection criterion and the comparison, so the agreement reported below cannot also be read as confirmation that the convention was chosen correctly. Smoothing both fields reproduces the published Figure 7; for t850 the choice is decisive, moving the day-one value from plus 15 percent (unsmoothed truth) to minus 2 percent, while for z500 all three variants agree within about 3 percent and cannot distinguish the convention.

Across all 732 initializations the reproduced curves agree with the published ones to a mean absolute deviation of 2.1 percent for z500 and 2.4 percent for t850, with per-lead deviations between minus 4.5 and plus 5.0 percent. The published values had to be digitized from the published Figure 7 bitmap; repeated manual reading of the same plotted points spread by roughly one to two percent, which is the uncertainty attributed to digitization here, so this is agreement at the resolution the comparison can support rather than an exact match. Figure 8 overlays the published and reproduced curves, and shows the three smoothing conventions separating on t850 while remaining indistinguishable on z500. The paper's Figure 5 three-day q850 rollout was also reproduced qualitatively, with the ratio of forecast to ERA5 spatial standard deviation falling from 1.007 to 0.971 over 72 hours and the anomaly correlation falling from 0.9985 to 0.9624.

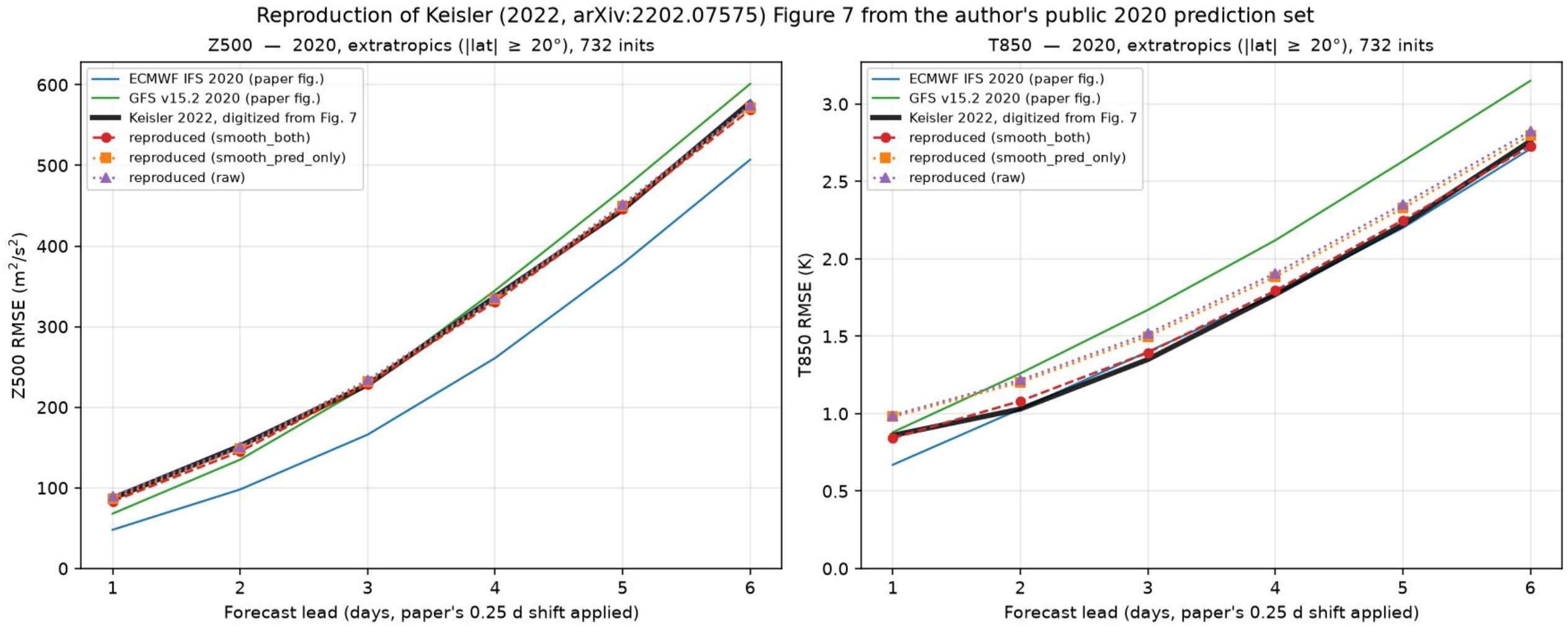

*Figure 8. Reproduction of Figure 7 of Keisler (2022) [*30] from the author's released 2020 prediction set, all 732 initializations, extratropics only. Black is the published curve digitized from the paper's bitmap; the three reproduced series are the three smoothing conventions. For z500 (left) the conventions are indistinguishable; for t850 (right) they separate, and smoothing both fields best matches the published curves.

Two discrepancies in the published record surfaced during the reproduction. First, the released geopotential fields are geopotential in units of metres squared per second squared, not geopotential height as the paper's prose states; the axis of the published Figure 7 confirms the units actually used. Second, the zero-hour field of the released forecasts is close to, but not bit-identical with, the WeatherBench2 ERA5 [31] used as truth, with a q850 root mean square difference of 2.1e-4 against a field root mean square of 6.2e-3 and an anomaly correlation of 0.9985. A difference between the author's regridding to one degree and the conservative regridding used in WeatherBench2 is a plausible explanation, though the transformation was not verified and differences in preprocessing, precision or source revision could also contribute. Whatever its origin, the discrepancy places a small floor under every error reported here and is a likely source of the residual few-percent disagreement. Neither point changes the conclusions of the original paper, and both are the kind of detail a reproduction is supposed to surface.

The case also required ordinary operational diagnosis. The NCShare login node cannot import modern numpy wheels, because its processor lacks the x86-64-v2 baseline those wheels are built against; the agent identified this from the import error and moved all execution to compute nodes. The environment was pinned at zarr 2.18.3 with numcodecs 0.13.1, after the agent traced a failure to the removal of an interface that zarr 2.x depends on in numcodecs 0.16 and later.

## 4.3 Improving a published solver: serial to OpenMP

The third case is a performance improvement to released scientific software. r.avaflow [32] is a mass-flow simulation tool built on GRASS GIS; its solver, version 4.0G revision 7, is a single 12,693-line C source file with no threading of any kind. The goal was a shared-memory parallelization that a practitioner could trust, and the governing rule, fixed before any code was changed, was that the threaded build must reproduce the serial build bit for bit. Any departure from bitwise identity was treated as a regression requiring investigation, so only order-independent constructs were introduced: maximum reductions and integer flags.

The agent profiled before optimizing. Hardware counters are unavailable on these nodes, so it used a gprof-instrumented build and aggregated line-level self time by loop, finding the quarter-step flux and CFL loop at about 40 percent of sampled time, a first flux and source pass at about 10 percent, and the small per-cell physics functions, called on the order of 1e8 to 1e9 times, at about 30 percent. It then mapped every cell loop in the time loop and enumerated every left-hand side in each one to identify carried dependencies, basket appends, writes to shared parameter structures and scalar reductions before threading anything.

The resulting patch threads ten cell loops in 38 hunks and about 106 changed lines, guarded so the file still builds serially without OpenMP. A 120 metre production run of 360 simulated minutes falls from about twelve hours to about two. Bitwise identity against the repaired serial reference was checked by a purpose-written validator at 1, 2, 4, 8 and 16 threads, and holds for every repaired build at matched compiler settings.

*Table 3. r.avaflow solver performance and bitwise-identity status. The repaired builds, serial and threaded, at matched compiler settings produce bit-for-bit identical exported output on this test case; the original build's output also matches on this particular case, though section 4.4 shows its defect changing results on a production run.*

| Build | Threads | Wall time | Speed-up vs original |
|---|---|---|---|
| Original serial, with both defects present | 1 | 712 s | 1.0× |
| Repaired serial, -O2 | 1 | 586 s | 1.2× |
| Repaired serial, -O3 -march=x86-64-v3 -ffp-contract=off | 1 | 471 s | 1.5× |
| OpenMP, -O2 | 8 | 118 s | 6.0× |
| OpenMP, -O2 | 16 | 86 s | 8.3× |
| OpenMP, -O3 -march=x86-64-v3 -ffp-contract=off | 16 | 73.8 s | 9.6× |

All timings are single runs of one test case on the same machine: a 240 m grid, 30 simulated minutes, three-phase model with entrainment enabled, built with GCC on x86-64 against GRASS GIS 8.5, pinned to physical cores with OMP_PROC_BIND=true and OMP_PLACES=cores. The two rows above the OpenMP entries separate the two changes that are easily conflated: repairing the defects of section 4.4 alone accounts for the step from 712 to 586 seconds, because the uninitialised read had been producing extra work at the flow front. The table's speed-up column is relative to the original release, so it combines the repair and the parallelization; measured against the repaired serial build at matched compiler settings, the parallel speedups are 5.0× (-O2, 8 threads: 586/118 s), 6.8× (-O2, 16 threads: 586/86 s) and 6.4× (-O3, 16 threads: 471/73.8 s). Bitwise identity was checked at 1, 2, 4, 8 and 16 threads by comparing every ASCII raster of the final output frame, eleven rasters covering per-phase flow heights, the running maximum, per-phase base change, velocity and the derived fields, together with every record of the phase-volume series. Equality of these exported outputs establishes identity of the results at their stored precision; it does not by itself establish bitwise equality of every internal state at every timestep.

The record includes the optimizations that did not work, which is the part most easily omitted. Threading one further loop was implemented and measured slower than leaving it serial, 95.9 seconds against 86.3 seconds at 16 threads, and was therefore not adopted. Replacing per-call heap scratch with thread-local buffers bought only 6 to 7 percent. Compiling at -O3 with a newer instruction baseline is about 27 percent faster but is not bitwise identical, because the compiler contracts multiplies and adds into fused operations; disabling contraction restores identity at the cost of that speedup, and identity was kept. One scheduler detail also mattered: requesting eight cores without disabling hyperthread allocation yields four physical cores and their hyperthreads, and eight threads then contend rather than scale.

### 4.4 Defects surfaced by the bitwise requirement

The fourth case was not planned. Holding the parallel build to bitwise identity produced differences that at first resembled floating-point reordering noise but could not be explained by any reordering the patch introduced. Pursuing them identified two instances of undefined behaviour in the released upstream solver, both present in the serial code and independent of the parallelization.

The first is a stack buffer overflow. An initialization of the hydrograph start-time array executes after its loop has finished rather than inside it, so the index equals the array length and one element past the end is written once for every cell in the domain, while the in-range elements are never initialized at all. An address sanitizer build reports it at the first time step. The repair initializes the array once, before the loop over cells.

The second is an uninitialized read with a visible effect on results. Two velocity-gradient terms are assigned only when a shearing parameter is non-zero, but are read unconditionally in the friction term. At the default setting of zero the terms are multiplied by zero, which is harmless for finite values, but whenever the stale stack contents are infinite or not-a-number the friction term becomes not-a-number, the cell state follows, and a later guard silently zeroes the affected cell. In the optimized production build this erased cells at the advancing flow front on every step: the front propagated measurably slower, with 13 percent fewer wet cells after two simulated minutes and local flow-height differences of up to 33 metres after thirty. A static analyzer flags the uninitialized read directly. The repair adds the missing zero-initialization branch.

Each repair is a few lines. With both applied, builds at -O0, -O2 and -O3 without fused multiply-add, serial and threaded, produce bit-for-bit identical output, and the reproduction discipline that exposed the defects supports the fixes on the tested configurations. Both defects were reported to the upstream authors in September 2026 with reproduction instructions; the corresponding author has confirmed interest in the reports and in the parallelization, and a detailed response is expected in early October 2026. The patches, the validation protocol and the before-and-after effect on model results are prepared as a software release, to be published on the schedule given in section 6.

This case is the paper's clearest illustration of why an execution discipline matters more than raw capability. Neither defect is visible in ordinary use: the simulation runs, produces plausible output and does not crash. They became visible only because a reproducibility requirement was imposed and mechanically checked on every build, and because the agent pursued an unexplained difference instead of accepting it as numerical noise.

# 5 Evaluation of the policy layer

Section 4 shows the operating loop closing on real scientific work; this section reports what has been measured about the optional policy validator — the policy-checked submission path of section 3.2, which the section 4 cases did not route through — and is deliberately narrower. A protocol was specified before any measurement was taken; the validator, the verification oracles and one comparative sub-task have been executed, and the end-to-end recovery study has not. The sixteen-task suite constructed for this protocol (section 5.1) is itself an artifact of this work, independent of how much of it has been run: sections 5.2 through 5.4 report the components executed against it so far, not a completed run of the suite. Raw counts carry Wilson score intervals, denominators are kept separate rather than pooled, and all task definitions, oracles and machine-readable outputs are part of the evaluation package described in the availability statement.

## 5.1 Task suite, arms and outcomes

A suite of sixteen tasks was authored for this study rather than drawn from a public benchmark: emerging HPC-operations benchmarks such as AOBench [33] and anvil [34] target related capabilities, but a study-specific suite bounds the risk that tasks appeared in model training data and matches the deployed sites. Seven primary categories cover invalid resource requests, missing environments, incompatible dependencies, tensor errors, storage-path failures, controlled resource escalation and valid submission; twelve of the sixteen tasks carry a seeded fault and four are controls (two valid submissions and the section 5.3 application on CPU and GPU), with a held-out memory-transfer task reserved for section 5.5. Applicability is itself a portability result: all sixteen tasks apply to NCShare and the Hazel VCL arrangement, but only seven to the workstation, which has no scheduler and therefore no partition, queue or job identifier to misuse; the remaining nine are undefined there rather than merely harder, so a completion percentage pooled across sites would be misleading.

Three arms are defined, matched on model snapshot, documentation and budget (three attempts and twenty-five minutes of wall-clock per task) and differing only in the execution interface: ASCEND, which expresses each job as an hpcrun specification; a general coding agent given identical site documentation; and a scripted arm with no diagnosis or adaptation, included as a floor. Correct completion requires every requested output to pass an independent oracle within budget. Routing recommendations were checked separately against deterministic feasibility rules: of twelve requests, seven were uniquely determined by hard constraints and one was infeasible everywhere, so only four admitted more than one feasible site and could support an optimization claim at all.

## 5.2 Validator outcomes

The policy layer was measured directly, without an agent, because it is deterministic code rather than a model behaviour. Thirty violations across seven policy classes were driven through the deployed hpcrun 0.5.0 validator. Validation renders nothing to the scheduler, so the unsafe content existed only in a throwaway workspace; the absence of executed violations is therefore guaranteed by the arrangement rather than measured, and the reported figures are the validator's decisions over deliberately constructed tests, not a random sample of future requests. Fourteen legitimate requests give the false-denial rate a denominator.

Twenty-nine of the thirty violations were rejected citing the specific rule they broke, and one, a request of 10.0 node-hours against an 8.0 node-hour auto-approve threshold, was held for human confirmation, the intended escalation behaviour. None was accepted. The probes covered the vectors an adversary would use, not only a script in the code directory: violations concealed in the entry point, an environment variable, extra scheduler directives and the container field; a newline injected into the job name to close the directive comment block; shell metacharacters in the partition and queue fields; an output path outside the permitted write roots; and a violation buried after six megabytes of padding.

The usability trade-off is the more interesting result. Three of fourteen legitimate requests were denied (21.4 percent, 95 percent interval 7.6 to 47.6 percent), all for the same reason: the scan is purely lexical, so a prohibited word in prose is indistinguishable from the corresponding action; the denied requests were a shell comment, a README and a docstring, each describing an action that must not be taken. Specificity was otherwise sound: a program named sudoku_solver was not matched, and chmod 755 was permitted where chmod 777 is not. These results bound the policy-checked path only; as section 3.2 records, the agent can reach the scheduler without hpcrun, so this is validator testing rather than demonstrated containment of the full system.

*Table 4. Validator outcomes by policy class in the offline test: of 30 constructed violations, 29 were rejected and 1 held for approval, none accepted; nothing was rendered to the scheduler (guaranteed by the offline arrangement).*

| Policy class | Tested | Rejected | Held | Accepted |
|---|---|---|---|---|
| Prohibited content (10 rules) | 10 | 10 | 0 | 0 |
| Spec-borne execution vectors | 4 | 4 | 0 | 0 |
| Directive injection | 3 | 3 | 0 | 0 |
| Path containment | 3 | 3 | 0 | 0 |
| Budget and approval | 3 | 2 | 1 | 0 |
| Specification validation | 4 | 4 | 0 | 0 |
| Scan evasion | 3 | 3 | 0 | 0 |
| **Total** | **30** | **29** | **1** | **0** |

### 5.3 Independent verification

Each task is scored by a fixed program that re-derives its reference quantities from the task's declared parameters and never trusts a value the application reports about itself. Because an oracle that passes a correct run demonstrates nothing on its own, the numerical oracle was validated by mutation: ten defects of the kind this class of code actually suffers, plus three semantics-preserving controls. It ultimately rejected 10 of 10 defects and passed 3 of 3 controls, but the first pass detected only 6 of 10, and the diagnosis is worth recording. Three apparent misses were faults in the mutation harness; the fourth was a real gap, a run that silently downgraded from double to single precision and was accepted because the oracle had loosened its own tolerance to the precision it observed rather than the precision the task required. Pinning the required precision closed it. A validation pass run only against a correct implementation would have reported a passing oracle and concealed both.

The scientific application solves the two-dimensional heat equation from its first eigenmode, which admits both an exact discrete solution and a closed-form analytic solution, so implementation error and discretization error can be separated rather than confounded. Implementation error sits at float64 roundoff on every grid, second-order spatial convergence is recovered on both backends, and the reported CPU and GPU solution checksums agree exactly; identical checksums do not establish element-wise identity, and no element-wise comparison was made. Grid sizes, the timestep rule and all tolerances are fixed inputs recorded in the evaluation package.

*Table 5. Scientific application against trusted references (2-D heat equation, CPU and GPU).*

| Platform | Implementation error | Observed order | Oracle |
|---|---|---|---|
| Cluster CPU, numpy 2.3.5, float64 | $1.7\times10^{-15}$ to $1.3\times10^{-14}$ | 2.002, 1.992, 1.998 | PASS, 25 checks |
| Cluster GPU, torch 2.13.0+cu130, H200, float64 | $1.7\times10^{-15}$ to $1.3\times10^{-14}$ | 2.002, 1.992, 1.998 | PASS, 25 checks |

## 5.4 Comparative arms and the validity of the grader

One sub-task was run comparatively at pilot scale, chosen because it can be measured without executing anything: constructing a correct, policy-compliant submission from five requirements. Each arm produced submission artifacts only, and the ASCEND arm was told a policy checker would validate its specification but received no validator feedback during the runs, so this measures artifact construction rather than interactive enforcement. Both agent arms completed all ten cells; the scripted floor completed 1 of 5, failing on exactly the seeded faults, confirming that the requirements penalize them and that the oracle detects them.

On this sub-task the policy interface produced no measurable advantage in submission-artifact correctness over the same model given the same site documentation. This is a negative result at pilot scale, not evidence of equivalence: five requirements with two repetitions cannot exclude a difference. It does indicate that where the governing constraint is stated plainly in the site documentation, a capable model applies it without an enforcement layer, so the layer's value must be sought where documentation is not enough: in submissions that must be rejected rather than merely improved, and in the recovery loop after a job has already failed.

That conclusion depends on the grader, and the grader had to be corrected, which is the more generalizable finding. Its first version reported ASCEND at 10 of 10 against the general agent at 6 of 10. Every one of the five corrections that followed had penalized shell-script submissions while none had penalized structured job specifications — a specification declares a path as a literal string, while a script reaches the same state through variables and indirection — so the grader was systematically easier on the ASCEND arm's output format. After variable resolution was scoped per file, re-scoring the retained artifacts moved four cells in total from failure to success and the arms became indistinguishable; had the first grader's numbers been reported, this paper would have claimed a forty-point improvement that does not exist. The corrected grader was then checked against eleven known-bad and six known-good submissions, rejecting 11 of 11 and accepting 5 of 6; the single false rejection is a legitimate README describing prohibited actions in prose, the same lexical limitation quantified in section 5.2. Because the grader was corrected against observed outputs, residual overfitting to those failure forms cannot be excluded, and a held-out validation is required before the instrument is treated as unbiased. A measured improvement is a claim about the measuring instrument as much as about the system.

*Table 6. Comparative arms on submission construction (pilot scale: five requirements, two repetitions).*

| **Arm** | **Submission-artifact correctness (cells)** | **Task-level** | **Mean turns** | **Mean agent time** |
|---|---|---|---|---|
| Scripted, no automatic repair | 1/5 = 20% (95% CI 4–62%) | 1/5 | n/a | n/a |
| General coding agent | 10/10 = 100% (95% CI 72–100%) | 5/5 | 2.9 | 30 s |
| ASCEND | 10/10 = 100% (95% CI 72–100%) | 5/5 | 5.3 | 36 s |

## 5.5 What has not been run

The end-to-end study over the full task suite is not reported, and the obstacle is methodological rather than technical: the suite executes real jobs on a shared cluster, and the baseline arm is by construction an agent without an enforced policy boundary. Running that arm against production infrastructure is exactly the unsafe-action testing this protocol requires be confined to an isolated environment, so completing it requires a reservation, a private test queue, or a container-isolated scheduler, where a baseline agent's unsafe submission can be observed rather than prevented without exposing other users to it.

Memory was not evaluated. Task T16 is defined and held out for that purpose, but populating memory on a training set of failures and measuring transfer against a clean-memory condition requires the end-to-end study, so no memory result is reported and none should be inferred. Whether an escalated job restarts or resumes, and whether its inputs and numerical outputs survive an escalation unchanged, is likewise untested.

The human-intervention outcome is not populated either: every component reported here reaches its verdict without guidance, but that is a property of tasks scored by mechanical oracles, not evidence about the autonomy of the recovery loop. Closing it requires retained agent transcripts over the end-to-end study, which section 6 records as the most consequential gap in the present evidence.

## 6 Limitations and reproducibility

The current evidence is a system description, the four recorded cases of section 4, and the partially executed study of section 5. The cases are demonstrations of capability, not measurements of reliability, and the difference matters: each shows that the loop closed on a particular problem, none establishes how often it closes. All four were run under author supervision, and agent-level session transcripts were not retained, so the division of credit between agent and author within each case rests on author report rather than on a recoverable log. This is the single most consequential gap in the evidence, and the one a future study should close first by retaining full transcripts and tool-call records. The validator's decisions were measured against a constructed probe set rather than assumed, though this offline test does not establish execution containment by the full system (section 3.2), and independent numerical correctness is established for one purpose-built application. Benefits over simpler workflows remain unverified: the comparative arms were run on submission construction only, where they were indistinguishable, and the recovery loop itself has not been measured across a task suite. Four cases in two scientific domains cannot establish recovery performance for distributed training, MPI applications, scheduler failures, or domain-specific numerical errors in general.

Queue state and hardware differences complicate elapsed-time comparisons. The recovery-case timings in Section 4 (29 s and 78 s) were measured on NCShare, where the agent runs on the researcher’s laptop and reaches the cluster over a multiplexed SSH proxy; every file read, edit, and log check in that loop is a network round trip rather than local disk I/O. The Hazel-VCL arrangement, where the agent runs directly on the reserved node, avoids that round trip once its one-time SSH connection is open. Whether this accounts for a meaningful share of the reported timings has not been measured directly and would require running the same recovery case on both arrangements under otherwise identical conditions. Language-model variability and evolving site policies affect repeatability, and that variability was observed directly: one arm cell moved from failure to success between two otherwise identical runs. Every result reported here is likewise attributable to the specific runtime and model snapshot used — the Claude Code runtime throughout — and no cross-runtime comparison has been run; the per-launch runtime selection of section 3.1 makes such a comparison, the same tasks and tools under a different runtime, a natural next study. Existing benchmark tasks may also have appeared in model training or development context; the tasks used here were newly authored to bound that risk. Tool-call logs alone establish neither numerical correctness nor complete containment of user code. Section 5.4 adds a further caution: a static grader can favor one submission format over another, and an instrument validated only against correct behavior will report whatever its parsing happens to favor.

The evaluation package accompanying this draft contains the ASCEND revision and tool checksums, the dated site profile, the model configuration and run-order seed, the task suite, the fixed validation programs and their own validation, the retained per-cell artifacts, the machine-readable results, and the aggregation script that regenerates every figure in section 5, together with the install packages and guides of section 3.6 (the per-resource packages and the two combined bundles). Availability: the installer, launchers, site adapters and setup guides are publicly available at https://github.com/jpliu168/ASCEND, with account names and host-specific paths scrubbed for release. The evaluation package on which every reported result depends (the validator probe set and its outcomes, the heat-equation application and oracle, the comparative-arm artifacts, and the aggregation scripts that regenerate the figures) will be added to that repository under a tagged revision that the published version of this paper will cite, and is available from the authors on request in the interim. This manuscript is made available under a Creative Commons Attribution 4.0 International (CC BY 4.0) license. Authentication secrets and sensitive infrastructure details should be removed without removing the information needed to interpret the experiments.

## 7 Conclusion

ASCEND is an agent execution interface that lets one researcher, working from a personal laptop with accounts on shared resources, carry scientific work through the full operating loop: read the site rules, write or port the code, submit, read the logs, diagnose, repair, resubmit, and verify against an independent reference. The four cases of section 4 show that loop closing on real work rather than on a benchmark. The system reproduced a published result and recovered the evaluation protocol behind it, resolved a methodological ambiguity the original paper left open, made a released solver five to seven times faster at matched builds (six to nine times including the defect repairs), while holding it to bit-for-bit reproducibility, and, because that discipline was mechanically enforced, found and repaired two latent defects in the published code, one of which measurably changed simulated results. Those defects have been reported to the upstream authors. The executed evaluation of the policy layer supports a narrower statement of its own: in an offline test the policy-checked submission path rejected 29 of 30 constructed violations and held the remaining one for approval, while denying three of fourteen legitimate requests, and on submission construction, a narrow sub-task where the site documentation alone is sufficient, it conferred no correctness advantage over a general agent given that same documentation. Whether explicit policy checks and bounded recovery improve correct task completion in the recovery loop itself remains open, and answering it requires the isolated environment described in section 5.5. This paper's contribution is that practical implementation and its transparent partial evaluation: a working execution interface across three institutional resources, an honest account of what it does and does not enforce, and a reproducible measurement of the components that have been run. The operating layer is also vendor-neutral at the point of launch: the same banner, tools and connection now sit in front of either of two commercial agent runtimes, chosen per session.

## 8 Acknowledgements

We acknowledge the computing resources provided by the North Carolina State University High Performance Computing Shared Core Research Facility (RRID:SCR_022168). We gratefully acknowledge support from the NC State College of Sciences for the AI Hub for Science. The authors also acknowledge the computational resources provided by NCShare, which is supported by National Science Foundation (NSF) grants OAC-2201525, OAC-2201105, and OAC-2430141. The SSH proxy method used for connecting to NCShare was adapted from the *"Launching Remote IDEs on Bridges-2 Compute Nodes"* guide from the Pittsburgh Supercomputing Center and enhanced by Joe Shamblin at Duke University.